\documentclass[twocolumn]{aastex7}

\newcommand{\mbh}{$M_{\rm BH}$}

\newcommand{\msun}{$M_{\odot}$}
\newcommand{\lsun}{$L_{\odot}$}

\usepackage{xcolor}

\shorttitle{MCG--06-30-15}
\shortauthors{Das et al.}

\DeclareUnicodeCharacter{0301}{\'}
\DeclareUnicodeCharacter{2212}{-}
\usepackage{mathtools}
\usepackage{soul}

\begin{document}

\title{A Stellar Dynamical Mass for the Central Black Hole in MCG--06-30-15}


\author[0000-0002-6870-6144]{Nabanita Das}
\affiliation{Department of Physics and Astronomy,
		 Georgia State University,
		 Atlanta, GA 30303, USA}
\email{ndas5@gsu.edu}

\author[0000-0002-2816-5398]{Misty C.\ Bentz}
\affiliation{Department of Physics and Astronomy,
		 Georgia State University,
		 Atlanta, GA 30303, USA}
\email{bentz@astro.gsu.edu}

\author[0000-0002-5038-9267]{Eugene Vasiliev}
\affiliation{University of Surrey,
        Guildford, GU2 7XH, UK}
\affiliation{Institute of Astronomy, 
        Madingley Road, Cambridge, CB3 0HA, UK}
\email{eugvas@protonmail.com}

\author[0000-0002-6257-2341]{Monica Valluri}
\affiliation{Department of Astronomy,
         University of Michigan,
         Ann Arbor, MI, 48104, USA}
\email{mvalluri@umich.edu}

\author[0000-0003-0017-349X]{Christopher A.\ Onken}
\affiliation{Research School of Astronomy and Astrophysics, 
        Australian National University, 
        Canberra, ACT 2611, Australia}
\email{christopher.onken@anu.edu.au}

\author[0000-0002-6248-398X]{Sandra. I.\ Raimundo}
\affiliation{Physics and Astronomy, University of Southampton, Highfield, Southampton, SO17 1BJ, UK}
\email{s.raimundo@soton.ac.uk}

\author[0000-0001-9191-9837]{Marianne Vestergaard}
\affiliation{DARK,  Niels Bohr Institute, University of Copenhagen, Jagtvej 155, DK-2200 Copenhagen}
\affiliation{Steward Observatory, University of Arizona, 933 N Cherry Avenue, Tucson, AZ 85721}
\email{mvester@nbi.ku.dk}

\begin{abstract}

We present the stellar dynamical mass of the central black hole in the nearby Seyfert galaxy MCG--06-30-15 using the Schwarzschild orbit-superposition method implemented in the open-source code FORSTAND. We obtained spatially resolved $K$-band nuclear stellar spectra for this galaxy with SINFONI on the VLT. We extracted the bulk stellar kinematics using Gauss--Hermite (GH) parameterization of the line-of-sight velocity distributions. A multicomponent surface brightness profile of the galaxy was determined from an \textit{HST} medium-band $V$ image. Our best-fit models indicate a black hole mass of $M_{BH}=(4.4\pm1.4) \times 10^7 M_{\odot}$ and a stellar mass-to-light ratio of $M/L$=($3.0\pm0.3$) \msun/\lsun, within 1$\sigma$ confidence intervals. Our constraint on $M_{BH}$ agrees with an upper limit on the mass from stellar dynamics based on the Jeans Anisotropic Method, but is $\sim$10 times larger than the reported mass from reverberation mapping. However, our best-fit $M_{BH}$ may be systematically biased high due to the counter-rotating disk in the nucleus of MCG--06-30-15 and the inability of the GH parameterization to fully describe such a complicated set of stellar kinematics. In addition, a dynamical $M_{BH}$ value depends heavily on the assumed source distance, which is not yet accurately constrained for this galaxy. MCG--06-30-15 is only the fourth galaxy in which we can compare $M_{BH}$ from stellar dynamical modeling with that from reverberation mapping. A direct comparison of $M_{BH}$ allows us to identify and investigate the possible sources of bias associated with different mass measurement techniques, which may influence our understanding of black hole and galaxy coevolution across cosmological timescales.
\end{abstract}


\section{Introduction} \label{sec:intro}
Supermassive black holes (SMBH), with masses ranging from approximately $10^6 M_{\odot}$ to $10^{10} M_{\odot}$, are found in the centers of massive galaxies. Black holes can be characterized by just three properties --- mass, spin, and charge --- whereas only mass and spin can be measured (although note the recent attempts to constrain charge by \citealt{ghosh23}). Therefore, black hole mass ($M_{BH}$) is a fundamental property that must be well constrained to accurately understand a black hole. Furthermore, the mass of a black hole determines the impact it may have on the surrounding environment: the nearby gas and its companion(s) in case of a stellar mass black hole (e.g., \citealt{valsecchi10}), and the host galaxy and circumgalactic environment in the case of a central SMBH (e.g., \citealt{ crenshaw12}).  Measurements of the masses of SMBH have also been found to correlate with a multitude of host-galaxy parameters such as luminosity, bulge stellar velocity dispersion, and bulge mass (e.g., \citealt{kormendy95,magorrian98,ferrarese00,gebhardt00,fabian12,kormendy13,bentz18}), which has been interpreted to mean that galaxies and their central black holes co-evolve.

To date, the most robust measurement of black hole mass has been made possible for Sagittarius A$^*$ (Sgr A$^*$) at the center of our own Galaxy by monitoring the individual stellar motions around it \citep{ghez00, genzel00, gravity22}. Although the size of the sphere of influence and the mass of the SMBHs at the centers of other galaxies are comparable to or larger than that of Sgr A$^*$, their vast distances make it impossible to resolve individual stars and measure their orbital velocities.

For central black holes in galaxies outside of our own, different approaches are necessary to measure $M_{BH}$. Some of the most widely accepted methods involve reverberation mapping --- measuring light echoes in photoionized gas when the black hole is accreting matter (e.g. \citealt{peterson93}) --- and dynamical modeling --- studying the spatially resolved kinematics of different tracers such as warm/cold gas, stars, or even water masers around the black hole (e.g. \citealt{vandermarel98}).

Reverberation mapping (RM, \citealt {blandford82, horne91, peterson93, cackett21}) is the most common $M_{BH}$ measurement technique for Type 1 active galactic nuclei (AGN), and is the basis for large numbers of AGN $M_{BH}$ estimates. RM can determine the structure and kinematics of the broad-line region (BLR) by measuring the time delay between variations in the continuum emission, which is produced by the accretion disk, and variations in the line emission from the BLR. The BLR gas is photoionized by the ionizing continuum, and at recombination, it emits spectral lines that we view as being broadened by the Doppler motions of the gas. Variability that occurs in the continuum flux (due to disk instabilities and/or fluctuating accretion rates) is reprocessed and echoed as variability of the flux from the BLR. One can measure the time delay between the variability features in the disk radiation and the response in the broad emission lines through spectrophotometric monitoring. This time delay is caused by the light-crossing time, and is therefore related to the characteristic size, of the BLR. When the time delay is combined with the BLR gas velocity, the mass of the black hole may be constrained. 

Since RM does not rely on spatial resolution but uses temporal resolution instead, this technique can be used to study both distant and nearby AGN. With very high-quality spectrophotometric monitoring data, velocity-resolved RM \citep{bentz10b, pancoast14b, grier17,brotherton20, bentz23b} may be carried out. Measuring the time delays as a function of the line-of-sight velocity across a broad emission line allows the kinematics and geometry of the broad-line region to be constrained, thus avoiding several assumptions and providing a direct and more accurate measurement of $M_{BH}$.

On the other hand, most other $M_{BH}$ measurement techniques require excellent spatial resolution. Water maser emission, for example, serves as an excellent probe of warm gas \citep{neufeld94} very close (~0.1$-$1 pc) to the supermassive black hole for nearby ($z \le$ 0.06) AGN \citep{rosenthal20}. By fitting Keplerian curves to the positions and velocities of the water maser emission in some systems, such as NGC 4258, the geometry and dynamics of gas in a thin circumnuclear disk may be constrained, thus providing a measure of $M_{BH}$ \citep{miyoshi95,herrnstein05, humphreys13, greene16}. However, there are not a large number of objects with observable maser emission \citep{Panessa20}. Additionally, nearly edge-on orientation of water masers in the accretion disk is a necessary condition to measure the mass of the central BH, so the technique is only suitable for Type 2 AGN (e.g., \citealt{zhang06,zhang10,greenhill08,castangia19}).

In gas dynamical (GD) modeling, several properties of the nuclear gas in a galaxy, such as the geometry and inclination (e.g., \citealt{macchetto97,denbrok15}) are deduced by spatially resolving the kinematics via observations of line emission arising from the gas.  Modeling of these constraints on the gas arrangement and dynamics allows the central BH mass to be determined. Cold gas usually serves as a better probe for mass measurements as it is found to be less turbulent (e.g., \citealt{barth16}). In the case of warm or ionised gas, which is more ubiquitous around AGN, non-gravitational perturbations (e.g., \citealt{Verdoes06}) and dust obscuration (e.g., \citealt{garcia15}) introduce additional challenges.

Stars, however, are not subject to turbulence. Stellar dynamical (SD) modeling measures $M_{BH}$ by constructing dynamical models of the bulk motions of the stars in a galactic nucleus (e.g., \citealt{kormendy95, vandermarel98, saglia16}). The dynamical models are fit to the observed line of sight velocity distributions (LOSVDs) of the stars and the surface brightness profile of the galaxy's stellar light. The stellar LOSVDs, which are obtained from spatially resolved spectroscopy, describe the bulk stellar kinematics as a function of spatial position. An accurate measurement of the distance to the galaxy is required to convert angular measurements into physical scales in the galaxy. In addition to an accurate distance measurement, the galaxy must also be fairly nearby so that the black hole sphere of influence (the region where the gravitational effect of the SMBH dominates over the gravity of stellar mass) is spatially resolved or nearly resolved \citep{davies06,gultekin11}. Gas dynamical modeling is also subject to this constraint, and so GD and SD modeling have only been applied to galaxies in the local neighborhood ($ D \lesssim 100 $ Mpc). We note that resolving the sphere of influence is also necessary in the case of the water maser technique, but radio arrays can reach higher spatial resolutions even at slightly larger source distances.
 
All of these black hole mass measurement techniques rely on several assumptions, and so comparing the results from multiple mass measurement techniques is necessary to reach better accuracy and precision. Our Galaxy is one of the few cases where two or more methods have been applied to measure $M_{BH}$ \citep{ghez00, genzel00, gravity22, eht19}. Additionally, the central black hole mass in M87 has been measured by SD modeling \citep{gebhardt11, liepold23} and modeling of EHT observations \citep{eht19}. Their derived masses agree with each other but are somewhat higher than the mass obtained from GD modeling \citep{walsh13}. The disagreement likely arises from the assumption of a pure Keplerian gas disk for M87 in the GD modeling analysis, where the black hole mass predicted by including non-Keplerian orbits \citep{Jeter19} would be similar to the other masses reported for this black hole.

There have been only a few comparisons of the masses from SD modeling and RM. Part of the reason is that the central AGN emits radiation that is necessary for implementing the RM method, but serves as a source of noise for SD modeling. Additionally, active galaxies in the local universe are rare, and the fact that they are primarily hosted in late-type galaxies makes them challenging to model. NGC 4151 \citep{bentz06b, onken14, derosa18, roberts21, bentz22}, NGC 3227 \citep{davies06, denney10, derosa18, bentz23b}, and NGC 5273 \citep{bentz14, merrell22} are the only three galaxies that have published comparisons of $M_{BH}$ measurements based on reverberation mapping and stellar dynamical modeling. While the masses derived from the two different methods are in agreement with each other for NGC 4151 and NGC 5273, the RM mass in the case of NGC 3227 was 4-5 times smaller than that from SD and GD until \citet{bentz23b} reported a new $M_{BH}$ from velocity-resolved RM.

MCG--06-30-15 is a nearby ($z = 0.00775$, \citealt{boisson02}) bright Seyfert 1 galaxy. The first broad Fe K$\alpha$ emission line with a relativistically redshifted tail was detected in this source \citep{tanaka95} and provided direct evidence of the presence of a spinning SMBH. Since then this AGN has been extensively studied due to its rapid X-ray variability (e.g., \citealt{arevalo05, mchardy05, chiang11, emman11, marinucci14, kara14}). \citet{mchardy05} performed long-term X-ray analysis at 0.1-10 keV and found a break frequency of $8^{+10}_{-3} \times 10^{-5}$ Hz in the power spectral density of the source. By assuming a linear relation between $M_{BH}$ and the break frequency for a sample of AGN with reverberation masses, they estimated $M_{BH} =  2.9^{+1.8}_{-1.6} \times 10^6 M_{\odot}$ for MCG--06-30-15. The first direct measurement of $M_{BH}$ was reported by \citet{bentz16a} using reverberation mapping, and they found  $M_{BH} = (1.6 \pm 0.4) \times 10^6 M_{\odot}$. \citet{raimundo13} constrained $M_{BH}$ using SD modeling based on the Jeans Anisotropic Modeling method \citep{cappellari08}, finding an upper limit of $M_{BH} \le  6 \times 10^7 M_{\odot}$. The spatial resolution of the stellar kinematics was too low to provide a stronger constraint.
In this paper, we report the highest spatial resolution observations of the nuclear stellar kinematics in MCG--06-30-15 to date. We describe the results of our work in applying stellar dynamical modeling using the \citet{schwarzschild79} orbit-superposition method, and we compare our results to the mass measurement derived from RM.

The content of this paper is presented as follows. In Section \ref{sec:obs}, we describe the observations that were acquired to constrain the spatially resolved stellar kinematics. Section \ref{sec:kinematics} describes the kinematic analysis, while in Section \ref{sec:Photometry} we present the details of the galaxy photometric constraints. Section \ref{sec: dynamics} focuses on the dynamical modeling using FORSTAND \citep{vasiliev20}. In Section \ref{sec:result}, we present our results while in Section \ref{sec: Discussion}, we discuss the interpretations and also compare our $M_{BH}$ with the previously determined RM mass. Finally, we summarize our findings in Section \ref{sec:summary}.

\section{Observations} \label{sec:obs}

MCG--06-30-15 is a Seyfert 1 galaxy at coordinates $\alpha = 13^h 35^m 53^s.70$ and $\delta = -34^{\circ} 17' 43\farcs9 $. The redshift of the galaxy is $z = 0.00775$. The estimated distance to the galaxy is $D=25.5\pm3.5$ Mpc \citep{tully13} based on the average distances of two galaxies in the same group as MCG--06-30-15.

Observations of MCG--06-30-15 (PI: Bentz) were obtained with the SINFONI integral field spectrograph \citep{eisenhauer03,bonnet04} on UT3 of the Very Large Telescope at the European Southern Observatory.  The $K$ grating and 100 mas/pix configuration were employed, giving a resolving power $R \approx 4000$ over the wavelength range $1.95-2.45$\,$\mu$m and a field of view of $3\farcs0 \times 3\farcs0$ with an effective spatial sampling of $0\farcs05 \times 0\farcs05$.  The observations were acquired in natural guide star mode, with the AGN serving as the guide ``star'', and with the instrument rotated to an on-sky position angle of $-30$\degr.  Over the course of seven nights between 2019 March and June, 46 observations were completed.  Typical exposure times were 300\,s, and observations were generally collected in groups of six with a standard star observed immediately after. 

The raw data were reduced with the EsoReflex \citep{freudling13} SINFONI pipeline v.3.3.0.  The pipeline carries out dark correction, flat fielding, wavelength calibration, spatial calibration, and rebuilds each individual observation as a data cube.  The individual cubes were then corrected for telluric absorption.  This process involved extracting a spectrum from one cube using a radius of 10 spaxels centered on the AGN, fitting the telluric absorption in the extracted spectrum with {\tt Molecfit} \citep{smette15,kausch15}, and then correcting the cube, spaxel by spaxel, for telluric absorption using the IRAF task {\tt telluric} and the template derived with {\tt Molecfit}.
After correcting for telluric absorption, each cube was shifted in velocity space to account for differences in the heliocentric velocity at the time of the observation. We rejected six observations that were taken under poor observing conditions, leaving us with 40 observations. Finally the cubes were aligned and combined with v.0.9 of the {\tt PyFu} package\footnote{http://drforum.gemini.edu/topic/pyfu-datacube-mosaicking-package/}, producing a final combined cube with an effective on-source exposure time of 3.33 hours.

\begin{deluxetable}{cccc}
\tabletypesize{\footnotesize}
\tablewidth{0pt}
 
 \tablecaption{PSF model \label{tab:psf}}
 \tablehead{
 \colhead{Component} & \colhead{Flux Contribution} & 
 \multicolumn{2}{c}{Width ($\sigma$)} \\
 \colhead{} & \colhead{(weight)} & \colhead{(arcsec)} & \colhead{(pc)}
  }
 
 \startdata 
        \hline
        1 & 0.12 & 0.04 & 4.9\\
        2 & 0.37 & 0.09 & 11.1\\
        3 & 0.51 & 0.24 & 29.5\\
        \hline
 \enddata
 \tablecomments{The relative flux contribution (weight) and Gaussian widths of the three components of the point spread function of our SINFONI data cube. The spatial scale of the components are calculated assuming a distance $D=25.5$ Mpc to the galaxy.}
\end{deluxetable}

An image of the point spread function (PSF) was created from the data cube. We began by taking several image slices from the continuum and several image slices from the wings of the broad emission line $Br_{\gamma}$ (rest wavelength = $2.166 \mu$m). We combined the image slices representing the broad-line emission and the slices representing the continuum separately, and then subtracted the continuum image from the broad-line image to isolate the spatially unresolved broad-line emission, which represents the PSF. We characterized this PSF using {\tt GALFIT} \citep{peng02,peng10}, an algorithm that can fit the surface brightness profiles of astronomical images with different functions. Table \ref{tab:psf} lists the parameters describing our best-fit model of the PSF that has three concentric and circular Gaussian profiles with different widths and weights.

\section{Kinematics}\label{sec:kinematics}
\begin{figure*}
    \centering
    \includegraphics[width=\textwidth]{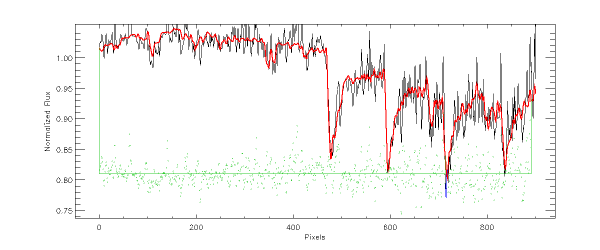}
    \caption{The spectral fit from \texttt{pPXF} for a selected bin at about $1''$ from the center. Data are shown in black, overplotted with the best-fit template in red, and the residuals (data $-$ model) are shown in green. Plotted on the X-axis are the pixel values corresponding to observed wavelengths from 2.19 to 2.42 $\mu $m while on the Y-axis, the corresponding fluxes normalized by the median value are shown. A noise spike appears at the wavelength associated with the blue line in several bins, and those pixels have been masked out during the fitting of all the spectra. The CO (2-0) absorption bands are the primary features we fitted to constrain the stellar line-of-sight velocity distributions.}
    \label{fig:specfit}
\end{figure*}

\begin{figure*}
    \centering
    \includegraphics[width=\textwidth]{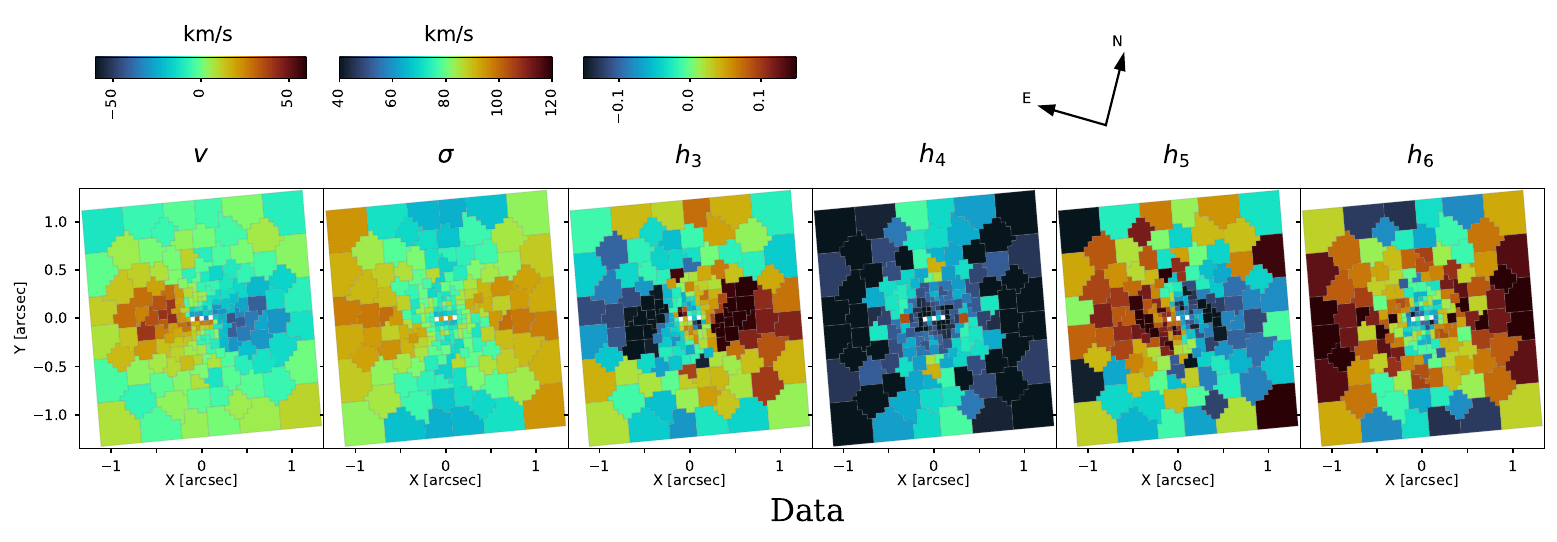}
    \caption{Kinematic maps derived from \texttt{pPXF} for the inner $3''\times3''$ (370 pc $\times$ 370 pc) of MCG--06-30-15 based on data we collected with SINFONI. The panels show the velocity ($v$), velocity dispersion ($\sigma$), and higher order ($h_3$-$h_6$) moments of the Gauss--Hermite polynomials for the data. The maps are rotated so that the kinematic major axis is parallel to the X-axis. The redshifts and blueshifts from the rotation of the nuclear stars are clearly visible in the velocity map. As shown in the color bars, red and blue correspond to higher and lower values, respectively. The higher order GH moments ($h_4-h_6$) use the same color scaling as $h_3$. The three white bins close to the center were severely affected by the AGN emission, and were masked out during the modeling. The data behind this figure is available in the machine-readable format in the online Journal.}
    \label{fig:kinmap}
\end{figure*}

With the fully reduced data cube, we began the kinematic analysis using \href{https://www-astro.physics.ox.ac.uk/~cappellari/software/}{\tt pPXF} (penalized pixel fitting, \citealt{cappellari04}). This software allows one to fit an input spectrum using a collection of stellar spectra. For our case, model galaxy spectra are constructed using combinations of stellar spectral templates from the NIFS $K$-band libraries (V1.5 and V2.0; \citealt{Winge09}), convolved with Gauss--Hermite approximations for the LOSVDs. The best-fit model spectra are found by minimizing the $\chi^2$ between model and observed galaxy spectra. To fit a spectrum, {\tt pPXF} requires a noise spectrum along with the actual galaxy data. The reduction pipeline of SINFONI does not produce any error spectra itself, hence synthetic noise spectra were created by combining the inverse of a telluric spectrum and the Poisson noise from the data in quadrature, to approximate the two most significant error sources: the background noise and the shot noise.

We also incorporated the Voronoi binning technique \citep{Cappellari03} to improve the signal-to-noise ratio (S/N), especially near the edges of the field of view (FOV) where the galaxy surface brightness decreases substantially. In this adaptive binning scheme, adjacent bins near the edge of the FOV are added together to increase the S/N while, close to the center, individual spaxels are assigned to their own bins. Thus, the spatial resolution is preserved in the region near the black hole.

To create the binning pattern, we divided the FOV in four quadrants where the kinematic major and minor axes act as the two axes of symmetry. The kinematic major axis is the line representing maximum rotation of the galaxy, and the line perpendicular to it is the kinematic minor axis. Using the method of \citet{Kraj06} and the stellar LOS velocities gleaned from the \texttt{pPXF} modeling, we determined the kinematic position angle in MCG--06-30-15 to be $74.5 \pm 24.8 ^{\circ}$ counterclockwise, relative to the y-axis of the SINFONI observations. Since the detector was rotated $-$30$^{\circ}$ from North when the observations were collected, the PA relative to North on the sky is $104.5 \pm 24.8 ^{\circ}$.

Symmetrizing the kinematics across the FOV is preferable for axisymmetric dynamical models. To carry out the symmetrizing, we determined the Voronoi binning in one of the quadrants, and reflected the same pattern into the three remaining quadrants.

The spectra associated with the spaxels within a single bin were coadded before fitting with \texttt{pPXF}. During the spectral fitting, we employed a point-symmetric technique where the spectra from two symmetric bins on each side of the kinematic minor axis are fitted by \texttt{pPXF} simultaneously. This improves the S/N and also ensures that the stars on either side of the kinematic minor axis have the same speeds, but move in opposite directions. Initial {\tt pPXF} fits for all bins were compared, and the model spectrum corresponding to the bin with the lowest $\chi^2$ was selected as the `optimal template'. This optimal template was then used to fit the data in all of the other bins. Selecting a single optimal template ensures that the stellar population remains constant in the FOV; loosening this assumption would allow the representative stellar population to potentially change sharply from spaxel to spaxel in an unphysical way.
Moreover, a fixed template ensures there are no differences from bin to bin because of stars with different metallicities, for example, contributing to the template and affecting the line shape measurements. Our best-fit template is comprised of five stars with spectral types: G8II, K2III, M5III, and M3III (2 of this type).

The wavelength range for the spectral fits extends from 2.19 to 2.42 $\mu$m as our goal was to primarily fit the strong CO absorption band heads near 2.29 $\mu$m, mostly found in the atmospheres of cool giant stars, while avoiding the strong emission lines from the AGN. Some of the bins in the FOV contain spectra affected by noise spikes between $2.371 - 2.373$ $\mu$m. We masked out these pixels in all the bins before fitting them. An example of a spectral fit is shown in Figure \ref{fig:specfit}, where the black and red lines denote the data and the template fit respectively, the green dots show the residuals after fitting the data with the best-fit template, and the blue lines represent masked pixels. 

The shapes of the CO absorption bands provide constraints on the LOSVDs of the stars. In \texttt{pPXF}, the LOSVDs are parameterized as a series of Gauss--Hermite (GH) polynomials of up to 6th order \citep{gerhard93, vandermarel93}. The first two moments ($h_1$ and $h_2$) of the GH polynomial are set to zero in the \texttt{pPXF} algorithm by default. Instead, the mean stellar velocity ($v$) and velocity dispersion ($\sigma$) are determined and reported along with the higher-order GH moments ($h_3-h_6$). The odd moments ($h_3$ and $h_5$) of the GH parameterization represent the degree of asymmetry of the fit from a pure Gaussian profile, similar to skewness, whereas the even moments ($h_4$ and $h_6$) represent the symmetric differences, similar to kurtosis.

The spectral fits can also be improved in {\tt pPXF} by changing the degree of additive or multiplicative Legendre polynomials. Additive polynomials reduce discrepancies from the template mismatch and background subtraction, whereas multiplicative polynomials help reduce the effects of reddening and calibration errors. As advocated in Section~2.4 of \citealt{cappellari09}, the use of additive polynomials is also beneficial for an accurate extraction of kinematics in the presence of AGN emission. In our case, we adopted additive and multiplicative polynomials of order 2 and 3, respectively, after testing several values and finding these were the smallest polynomial orders needed to minimize  low-frequency  mismatches in the continuum shape between the templates and the observations. The penalty term in pPXF penalizes the fit for noisy spaxels by forcing higher ($n\geq 3$) order moments of the GH polynomials to be close to zero, while retaining the higher-order moments in fits for spaxels with low noise. We tested penalty values from 0 to 0.5 in steps of 0.1 and based on these tests we adopted a penalty value of 0.3 in our best-fit model.

Our final kinematic maps from \texttt{pPXF} are shown in Figure \ref{fig:kinmap}. The maps are rotated to show the maximum velocity along the horizontal axis. Three of the central bins, shown in white in Figure \ref{fig:kinmap}, were severely affected by the AGN light. These bins were discarded because the LOSVDs from them were not reliable for dynamical modeling.

\section{Photometry}\label{sec:Photometry}
While the spatially resolved spectra provide constraints on the kinematics, photometric analysis is required to constrain the luminosity distribution of the galaxy, or the surface brightness profile. In addition, reconstructing the 3D stellar mass distribution also requires initial estimates of the mass-to-light ratio as well as the inclination of the galaxy to our line of sight.

\subsection{Surface Brightness Decomposition} 
We used {\tt GALFIT} \citep{peng02, peng10} to build a multicomponent surface brightness profile for MCG--06-30-15. Based on \textit{HST} WFC3 imaging through the F547M filter from \cite{bentz16a}, our model for MCG--06-30-15 consists of 3 components for the galaxy: bulge, disk, and a faint round extended component encompassing the disk and the bulge. The AGN was fit with a model PSF derived from the \texttt{Starfit} algorithm \citep{hamilton14}, and a gradient for the sky background was included. The best-fit parameters for each galaxy component are listed in Table \ref{tab:sb}: integrated apparent magnitude in the $V$ band ($m_V$), effective radius ($R_e$), S$\text{\'e}$rsic index ($n$), axis ratio ($q^{obs}$), and position angle (PA).  We adopted an offset of $V - F547M =  -0.0259$\,mag for the color correction and a value of $A_V = 0.165$\,mag to account for the dust extinction in our Galaxy along the line of sight \citep{schlegel98,schlafly11}. These corrections are included in the magnitudes reported in Table \ref{tab:sb}. During the fitting, the PA and the centers were tied together for all the components of the galaxy as a requirement for our dynamical modeling code. The reported PA ($-32.2\degr$) is measured from the Y axis of the image, which was rotated by $-32.84 \degr$ from the North. Therefore, the photometric PA for the galaxy disk is $\simeq -65 \degr$, or $115 \degr$ ($180 + (-65) \degr$). The dust lane near the center of MCG--06-30-15 was masked out before modeling the surface brightness profile.

\begin{figure}
    \includegraphics[width=3.3 in]{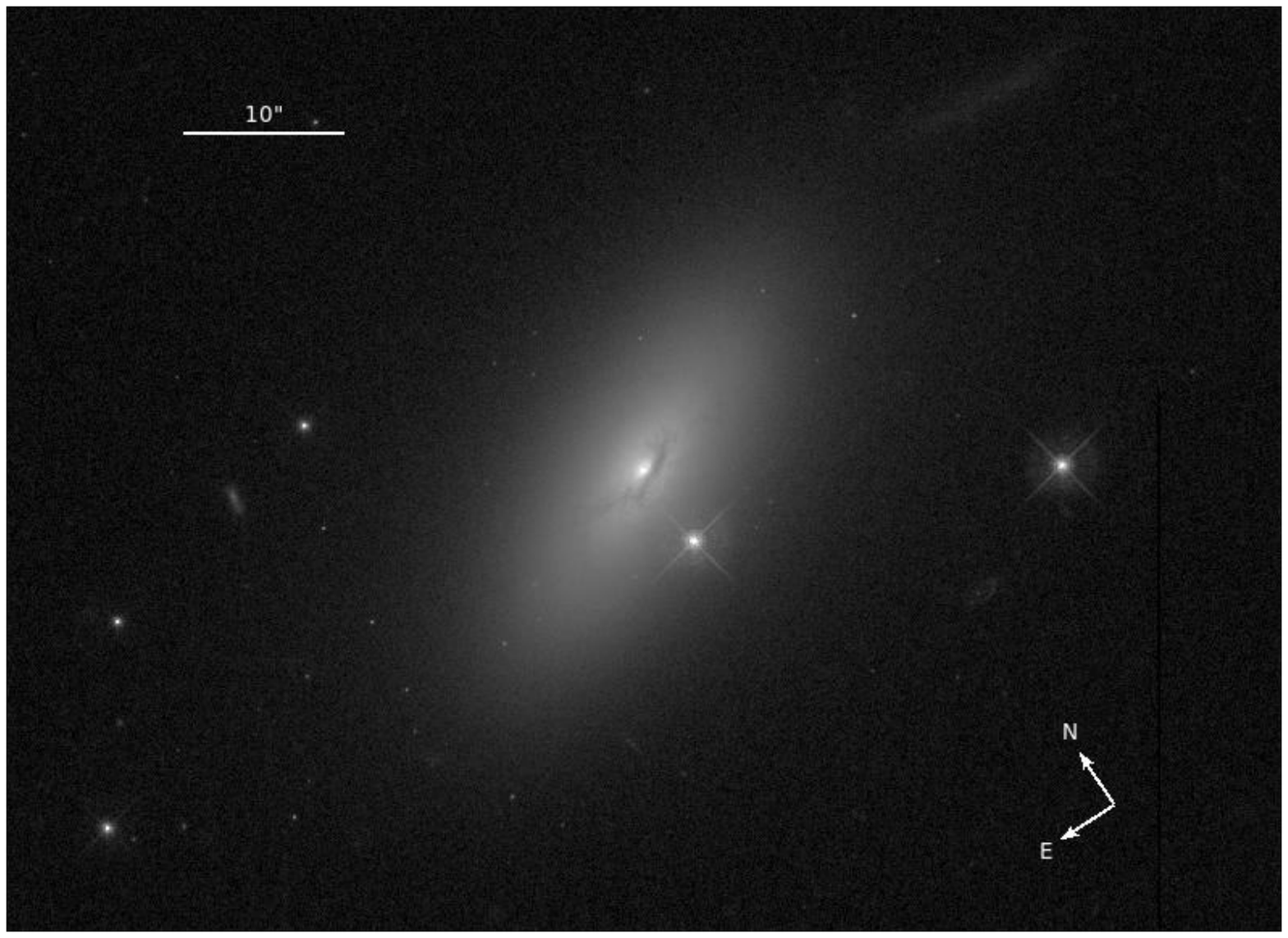}
    \begin{minipage} {0.45\textwidth}
         \caption{\textit{HST} WFC3 F547M (medium-band $V$) image of MCG--06-30-15 \citep{bentz16a}. The scale for $10''$ is shown. The dust lane in the galaxy disk was masked before fitting the surface brightness profile.}\label{fig:mcg06_hst}
    \end{minipage}
\end{figure}

\begin{deluxetable}{cccccc}
\tabletypesize{\footnotesize}
\tablewidth{0pt}

\tablecaption{ \texttt{GALFIT} parameters for Surface Brightness \label{tab:sb}} 
\tablehead{
\colhead{Components} & \colhead{$m_V$} & \colhead{$R_e$ } & \colhead{$n$}& \colhead{$q^\text{obs}$} & \colhead{PA}\\
 \colhead{} & \colhead{(mag)}  & \colhead{(arcsec)}  & \colhead{}  & \colhead{} & \colhead{ (degree)}
  }

 \startdata
        Bulge & 15.62 & 1.16 & 2.00 & 0.69 & $-32.20$\\
        Disk & 13.78 & 7.76 & 0.93 & 0.41 & $-32.20$\\
        Stellar halo & 14.36 & 10.74 & 1.19 & 0.78 & $-32.20$
 \enddata
\tablecomments{The photometric components of the galaxy and their parameters: $V$-band integrated magnitude ($m_V$), effective radius ($R_e$), S$\text{\'e}$rsic index ($n$), axis ratio ($q^\text{obs}$), and position angle (PA) measured from the Y axis of the image respectively. The position angle for all components was required to be the same for dynamical modeling.}
\end{deluxetable}

\subsection{Mass-to-Light Ratio}
\label{surface}
The stellar mass-to-light ratio ($M/L$, or $\Upsilon$) of a galaxy indicates the mass present in the galaxy from luminous stars. We estimated the stellar $M/L$ of the galaxy from its color using Table 1 of \citet{bell01}. After carefully removing the AGN contribution from the \textit{HST} WFC3 F547M image (Figure \ref{fig:mcg06_hst}, \citealt{bentz16a}) and from a $K$-band image from the VHS survey \citep{mcmahon13}, we determined $V - K = 3.1 \pm 0.2$ mag for the galaxy, and estimate the $V$-band $M/L$ ratio, $\Upsilon_{V, V-K} = (4 \pm 1) $ \msun/\lsun.

\subsection{Galaxy Inclination}
The inclination angle is an important component for accurately modeling the galaxy because images only provide a 2D projection of the 3D galaxy. The inclination angle also determines the component of the velocity vector that is visible along our line of sight. The projected shape of the disk of the galaxy may be modeled using concentric ellipses. The inclination angle to our line of sight can be estimated from the apparent major-to-minor axis ratio of the disk

\begin{equation}\label{eq:incl}
    \cos (i) = \left[{\frac{{q_d^\text{obs}}^2  - {q_{0,d}}^2}{1 - {q_{0,d}}^2}}\right]^{0.5}
\end{equation}

\noindent as shown by \citet{hubble1925}. In Equation \ref{eq:incl}, $q_d^\text{obs} = b/a$ is the observed disk axis ratio and $q_{0,d}$  or the global flattening parameter, is the axis ratio of a perfectly edge-on disk galaxy. Following \citet{tully00}, we adopt $q_{0,d} = 0.2$, but explore the consequences of varying this parameter in Section~\ref{sec:params_varying}. The best-fit disk axis ratio from the surface brightness fitting gives $q_d^\text{obs}=0.41$ for MCG--06-30-15 (see Table \ref{tab:sb}), which corresponds to an inclination angle of $69^{\circ} $ between the disk of the galaxy and our line-of-sight. 
\section{Dynamical Modeling}\label{sec: dynamics}

To construct the stellar dynamical models, we have used FORSTAND \citep{vasiliev20}, publicly available within the \href{https://github.com/GalacticDynamics-Oxford/Agama}{AGAMA} library \citep{vasiliev19}. This code implements the \citet{schwarzschild79} orbit-superposition method to create dynamical models. For each model, first, a deprojected 3D stellar density profile is determined from a surface brightness fit derived from an image. Then the routine computes the corresponding gravitational potential including other input parameters, e.g., mass-to-light ratio ($\Upsilon$), contribution of dark matter (DM) halo, initial guess for $M_{BH}$, etc. An orbit library is constructed with 20000 random stellar orbits that are integrated within each trial of gravitational potential. The code then determines a weighted superposition of orbits to match the observed kinematic constraints and records the goodness of fit for the model. The best models are characterized by minimizing the $\chi^2$ statistic. 

We summarize below the important inputs for the models as two separate groups: 1.\ non-varying inputs, where a single value is adopted for all model runs,
 and 2.\ varying parameters, where different potential values are explored in the model runs.

\subsection{Non-varying (single value) inputs}  \label{sec:params_fixed}
\begin{itemize}

    \item \textbf{Kinematic constraints:} The observed kinematics as described in Section \ref{sec:kinematics} determine the number of degrees of freedom or kinematic constraints for the models, and constrain the total mass. Since \texttt{pPXF} employs a point-symmetric fitting to extract the LOSVDs of the stars, we only used half of the kinematics to construct the models. The number of kinematic constraints is determined by multiplying the number of GH moments in the LOSVDs (here, six) by the number of independently fitted Voronoi bins, which is half of all the bins (99 in this case) minus the number of fitted parameters (here, two: $M_{BH}$ and $M/L$). Our dataset for MCG--06-30-15 contains 582 kinematic constraints.
  
    \item \textbf{Surface brightness profile:} Based on the results from \texttt{GALFIT} described in Section \ref{surface}, the surface brightness profile for MCG--06-30-15 consists of three components: an ellipsoidal bulge, a flat disk, and a faint outer spheroidal component. The integrated magnitude in $V$ ($m_V$), effective radius ($R_e$), S$\text{\'e}$rsic index ($n$), axis ratio ($q^{obs}$), and position angle (PA) for each component are tabulated in Table \ref{tab:sb}. The surface brightness profile parameterizes the flux contribution from each component of the galaxy as a function of location in the galaxy, which estimates the stellar mass profile for the dynamical models when combined with the stellar $M/L$.
    
    \item \textbf{Distance:} The distance to the galaxy is required to convert angular sizes to physical sizes. If $v$ is the measured velocity within distance $r$ from the black hole, then the black hole mass $M_{BH}$ is related to $r$ by the following expression: 
    \begin{equation}
        M_{BH} \propto \frac{v^2 r}{G}
    \end{equation}
    where $G$ is the gravitational constant. We measure angular sizes on the sky, which are directly proportional to physical sizes modulo the assumed distance $D$. Therefore, a galaxy that is more distant than assumed would have a larger black hole mass than reported by the models. For MCG--06-30-15, we adopted a distance of $D = 25.5$ $\pm$ 3.5 Mpc \citep{tully13}, which is so far the best estimate but is not very reliable (see Section \ref{sec: Discussion} for additional discussion of the distance).
    \item \textbf{Point Spread Function:} The point spread function convolved with the LOSVDs corrects for the effect of the finite spatial resolution of the spectrograph, in this case SINFONI. FORSTAND requires two pieces of information to describe the PSF: the width and flux contribution(s) of either a single Gaussian or several Gaussian components summed together. In the latter case, the flux contributions should sum up to 1. We used the PSF model reported in Table \ref{tab:psf} as our input. 
    
    \item \textbf{Number of orbits:} We integrated 20,000 random stellar orbits within the gravitational potential for our different model runs. To obtain the best fit for the kinematic data, the ratio of the number of orbits to the number of kinematic constraints should be $\gtrsim 5$ \citep{valluri04}. In case of a low number of orbits, the models may be overconstrained and yield unreliable results for the free parameters.
\end{itemize}  
\subsection{Varying parameters}  \label{sec:params_varying}
\begin{itemize}

    \item \textbf{Stellar mass-to-light ratio:} We estimated an initial value for the stellar $M/L$ based on the observed color of the galaxy (discussed in Section \ref{surface}), and found the $V$-band mass-to-light ratio $\Upsilon_{V, V-K} = 4 \pm 1 $ \msun/\lsun. The code adopts this initial value of $M/L$ for each model, and constructs more models by varying the initial $M/L$ in steps of $2^{0.05}$ (or another value selected by user).

    \item \textbf{Black hole mass ($M_{BH}$):} We started with $M_{BH}$ values from $0 - 5 \times 10^7 M_{\odot}$ in steps of $\Delta M_{BH} = 2.5 \times 10^6 M_{\odot}$ as inputs. This choice allowed us to optimize the computational time required to scan a large range of possible mass values. Every $M_{BH}$ input is used to generate a list of models for all possible $M/L$ values. In other words, a grid of models in the $M_{BH}$ vs.\ $M/L$ space is created by varying the initial $\Upsilon$ by a multiplicative factor of $2^{0.05}$, and at the same time multiplying the input $M_{BH}$ by the corresponding $\Upsilon$. The best value for $M_{BH}$ is determined from the $\chi^2$ surface that is generated from the grid of models. Examination of our initial results suggested that a finer grid of $M_{BH}$ values was not necessary.
    
    \item \textbf{Flattening of galaxy components $-$ disk ($q_d$), bulge ($q_b$), stellar halo ($q_s$):} We explored a range of values for the flattening of the different components of the galaxy. The photometric analysis from \texttt{GALFIT} for MCG--06-30-15 has three different components: one flat disk-like component, one slightly spheroidal bulge, and a faint, round, extended structure around the disk and the bulge, which we refer to as the ``stellar halo'' below. We note that the disk flattening value ($q_d$) can be computed using the observed axis ratio ($q_d^\text{obs}$) from \texttt{GALFIT} and Equation \ref{eq:incl}. Instead, we treated $q_d$ as a free parameter, and estimated inclination from it. Similarly, we examined a range of flattening values for the bulge and the stellar halo instead of just using the \texttt{GALFIT} axes ratios to effectively explore the consequences of possible uncertainties in these values. We constructed models for disk flattening ($q_d$) of 0.1, 0.2, and 0.3. Each of these $q_d$ values was used in the construction of models with different flattening for the bulge and the stellar halo, where $q_b$ and $q_s$ were explored over the range of $0.2-1.0$ with steps of $0.1$.

   \item \textbf{Dark matter halo:} The dark matter (DM) halo is the most massive component of a galaxy, but its potential dominates over the baryons primarily in the outer parts of a galaxy (e.g., \citealt{rubin62, rubin78}), which we do not model due to the limited radial range of our kinematic data. FORSTAND allows users to set the rotational velocity ($V_{DM}$) and scale radius ($r_{DM}$) of the dark matter halo to explore the impact on the modeling results from different assumptions about the DM halo properties. We tested models with and without the presence of a DM halo. For the models including dark matter, we used the Navarro–Frenk–White profile \citep{navarro96} with a scale radius of $r_{DM}=180''$ ($22.5$ kpc) and DM halo circular velocity, $V_{DM}=100\sqrt{\Upsilon}$ km/s. This value of DM halo circular velocity is similar to what has been found for Milky Way-like galaxies \citep{deisidio24}. We estimated $r_{DM}$ using the stellar mass -- halo mass relation of \citet{behroozi19}, and the halo mass -- concentration relation of \citet{dutton14}. 

    \item \textbf{Random seed:} The random seed parameter in FORSTAND is responsible for setting the initial conditions for the generated orbit library. We explored different seed values to test different randomized initial conditions for our best-fit model. We note that the results reported here are not significantly affected by changing the seed value, indicating our modeling results are robust and stable.

\end{itemize}  
\section{Results} \label{sec:result}
\begin{figure*}
    \centering
    \includegraphics[width=\textwidth]{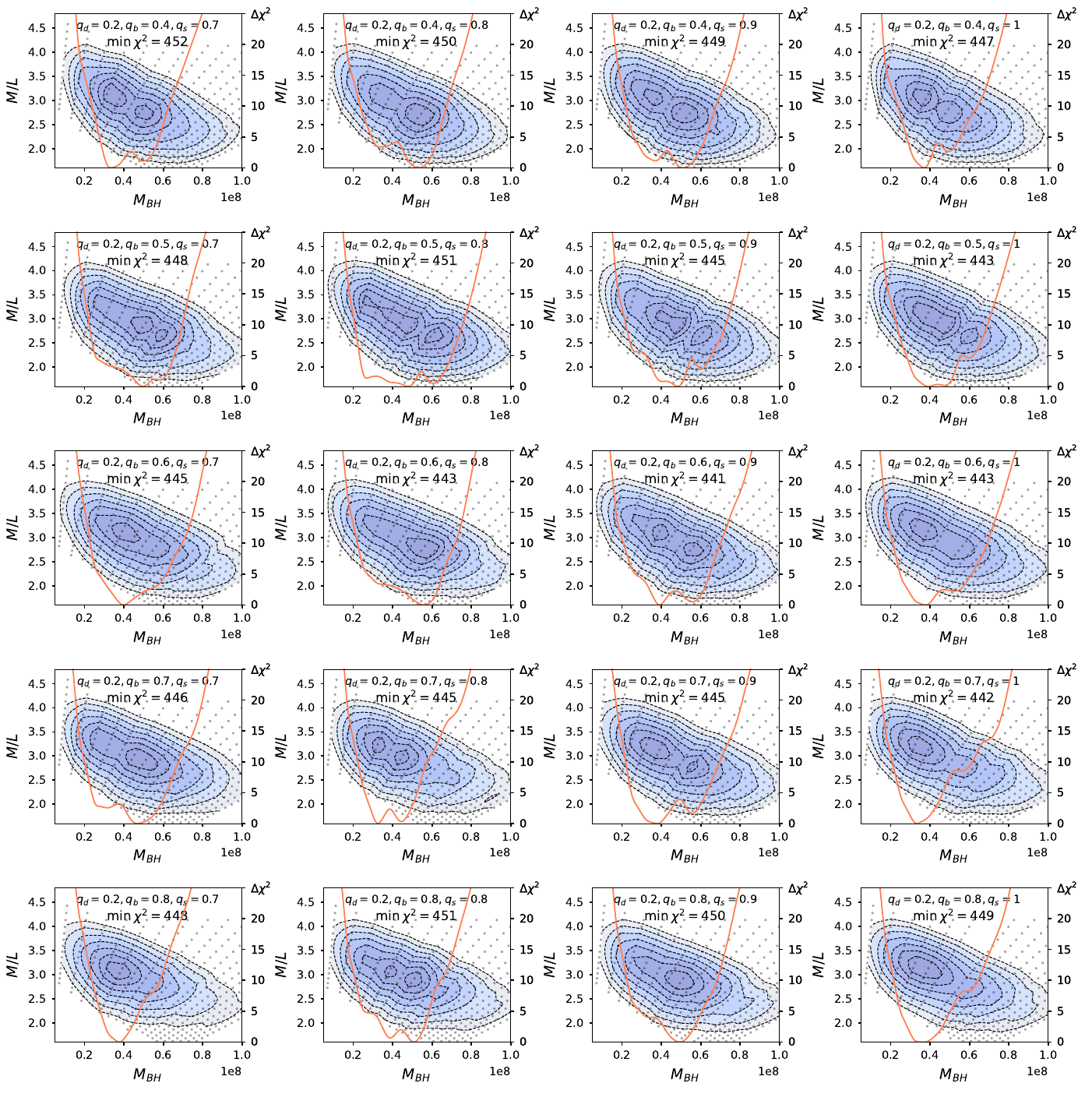}
    \caption{Comparison of $\Delta\chi^2 (\equiv \chi^2 - \chi^2_{min})$ as a function of $M_{BH}$ and $M/L$ for dynamical models with different galactic geometries. The dashed contour lines along with the blue density profiles indicate the 1$\sigma$, 2$\sigma$, 3$\sigma$, ... confidence levels of 2D $\Delta\chi^2$ at 2.3, 6.2, 11.8, ... etc. The orange lines show the $\Delta\chi^2$ value as a function of $M_{BH}$ marginalized over $M/L$. The intrinsic flattening of the bulge ($q_b$) varies down the columns and is fixed across the rows, whereas the intrinsic flattening of the stellar halo ($q_s$) varies across the rows, and is constant down the columns. The intrinsic flattening of the disk ($q_d$) is held fixed at 0.2 for all of the panels. The models with the lowest $\chi^2$ are found in the third and fourth rows with $q_b= [0.6-0.7]$ and $q_s = [0.7- 1.0]$, and in the second row with $q_b = 0.5$, and $q_s = [0.9-1.0]$.}
    \label{fig:results}
\end{figure*}

\begin{figure}
    \includegraphics[width=3.1in]{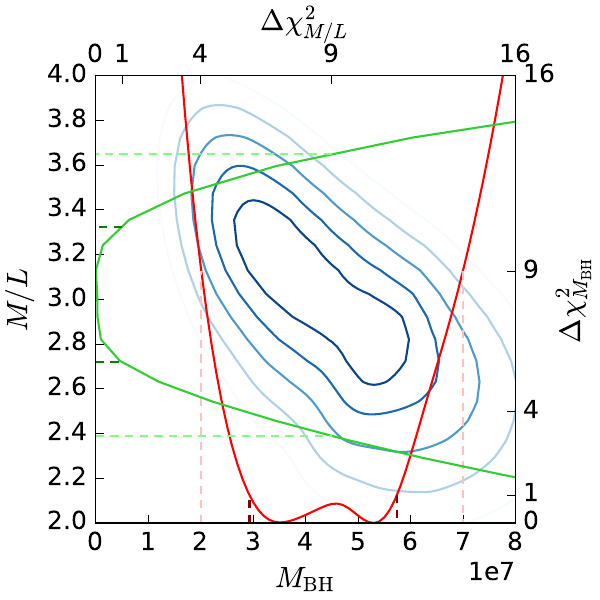}
    \begin{minipage}{0.44\textwidth}
         \caption{1D and 2D profiles of $\Delta\chi^2$ in the $M_{BH}$ and $M/L$ parameter space, marginalized over $q_b$ and $q_s$ (the other two parameters shown in Figure \ref{fig:results}), as explained in the text. Blue contours have the same meaning as in the previous figure, while red and green curves show the marginalized 1D profiles for $M_{\rm BH}$ and $M/L$ respectively. Dark and light dashed lines show the 1$\sigma$ and 3$\sigma$ confidence intervals on each of these two parameters.}\label{fig:chi2_mbh}
    \end{minipage}
\end{figure}

Since the models are dependent on a large number of parameters as discussed above, we determined the best-fit models by systematically changing the parameters one by one. Our free parameters are: black hole mass ($M_{BH}$), stellar mass-to-light ratio ($M/L$), flattening of the bulge, disk, and stellar halo ($q_b$, $q_d$, and $q_s$), and DM halo contribution. 

We first explored different shapes for the galaxy by varying the flattening of the components in the absence of a DM halo. A grid of the constructed models is shown in Figure \ref{fig:results}. The figure shows a comparison of $\Delta\chi^2$ =$\chi^2 - \chi^2_{min}$ as a function of $M_{BH}$ on the horizontal axis and stellar mass-to-light ratio on the vertical axis. Each panel represents a single galaxy shape, and the grey dots denote the models that were generated. The dashed contours and blue density profiles in each panel represent the 1$\sigma$, 2$\sigma$, 3$\sigma$, .... confidence levels of $\Delta\chi^2$ for the two parameters, $M_{BH}$ and $M/L$, and the orange curves show the  $\Delta\chi^2$ as a function of $M_{BH}$ marginalized over mass-to-light ratio. The bulge flattening ($q_b$) is kept constant along the rows and varies down the columns, while the stellar halo flattening ($q_s$) is held fixed along the columns and changes across the rows. The disk flattening is kept constant for all the panels at $q_d = 0.2$, which is a common value measured from edge-on disk galaxies (e.g. \citealt{hubble1925,tully00,kautsch06}), though we also explored these combinations of parameters for $q_d=0.1$ and $q_d=0.3$.
The difference in lowest $\chi^2$ values is statistically insignificant for several of the models in the middle rows of Figure \ref{fig:results} with $q_b= [0.6-0.7]$ and $q_s= [0.7-1.0]$, and for $q_b = 0.5$ with a round halo, $q_s = [0.9-1.0]$. In the following discussion, we refer to these as our best-fit models. The range for $\chi^2$ minima in these best-fit models is between $441-446$, while the reduced $\chi^2$ values range between $0.76-0.77$.

In Figure~\ref{fig:chi2_mbh}, we condense the above 4D grid of best-fit models into 2D contours of $\Delta\chi^2$ as functions of $M_{\rm BH}$ and $M/L$, as well as 1D profiles in each of these parameters. These values are obtained by marginalizing over the nuisance parameters $q_b$ and $q_s$. Namely, first we interpolate the $\chi^2$ values onto a regular grid in the 4D space ($M_{\rm BH}, M/L, q_b, q_s$), using either radial basis function or gaussian process regression interpolators (the results are very similar); then we convert these values into likelihoods $\mathcal L = \exp(-\chi^2/2)$ and integrate them over the nuisance dimensions; and finally we convert likelihoods back into $\chi^2$ values. Because the locations of the minima of $\chi^2$ vary across panels in Figure~\ref{fig:results}, but their absolute values are quite similar, the marginalized 2D profile has the shape of a rather flat trough, and the marginalized 1D curves have flat-bottomed profiles. 
The $n$-$\sigma$ confidence intervals on each parameter are obtained from these 1D profiles as $\Delta\chi^2=n^2$. Specifically, the 1$\sigma$ (68\%) range for $M_{\rm BH}$ is [3.0--5.8]$\times 10^7$\,\msun and for $M/L$ is [2.7--3.3]\,\msun/\lsun. Because the contours are not parabolic, the 3$\sigma$ (99.7\%) intervals are less than three times broader than the 1$\sigma$ intervals.

From the best-fit models, we have selected the model with $q_b = $ 0.6 and $q_s = $ 0.8 as our `fiducial model' --- a representative model that we use to demonstrate how the results change when we adjust one parameter or another. In Figure \ref{fig:kin_comp}, we show the observed kinematics as well as the kinematics of the fiducial model and the kinematic residuals between them. The residuals are defined as data minus model, normalized by the observed errors ($\epsilon_{obs}$) in the GH moments.

Figure \ref{fig:chi2_dis} shows how the results from the dynamical models depend on other parameters. In the top panel of Figure \ref{fig:chi2_dis}, we show the dependency of $M_{BH}$ on the distance to the galaxy. As discussed in Section \ref{sec: dynamics}, our models show that $M_{BH}$ scales linearly with the distance, yielding smaller $M_{BH}$ if the host galaxy is nearer, and larger $M_{BH}$ if the galaxy is farther than assumed here.

We also explored different values for the disk flattening $q_d$, where the disk flattening relates directly to the inclination angle at which we are viewing the galaxy. A comparison between different disk flattening values is presented in the middle row of Figure \ref{fig:chi2_dis}. The $\Delta\chi^2$ value increases for models with $q_d$ = 0.1, but models with $q_d$ = 0.3 have lower $\Delta\chi^2$ values. However, the best-fit $M_{BH}$ and $M/L$ values do not change significantly for different disk flattening values.

We have also tested the effect of DM in our models and find that the results are nearly identical when DM is included or excluded. A comparison of models including and excluding the effect of a DM halo is shown in the bottom row of Figure \ref{fig:chi2_dis}, and we see negligible difference.
The FOV ($3''$ $\approx$ $370$ pc) of our data cube is small enough that the stellar kinematics seem to be largely unaffected by the DM halo.

\section{Discussion} \label{sec: Discussion}

\subsection{Mass-to-Light Ratio}
Based on the galaxy color, we initially estimated a $V$-band mass-to-light ratio of $M/L$ = ($4 \pm 1$) $M_{\odot}/L_{\odot}$. Inferring the $M/L$ from the color of a galaxy is not always precise as it depends on many factors such as the star formation history, metallicity, etc \citep{haghi17}. Nevertheless, our best-fit models from FORSTAND agree with this simple estimate, finding that the $V$-band $M/L$ is $(3.0\pm0.3)$  $M_{\odot}/L_{\odot}$.

We note that the surface brightness profile of MCG--06-30-15 is affected by a dense dust lane in the disk, which is nearly edge-on. For the surface brightness modeling, our approach was to use a medium-$V$ \textit{HST} image \citep{bentz16a} and to carefully mask the pixels affected by dust. We explored the use of an $H$-band \textit{HST} image for creating the surface brightness parameterization, but we found that while the dust extinction was lessened, it was not entirely absent. Furthermore, the central region, where the gravity from the SMBH dominates, is not as well resolved in the $H$ band as it is in the $V$ band due to the three times poorer spatial resolution. Ideally, one should create a dust profile to accurately account for the extinction. However, \citet{boizelle19} have argued that the creation of a spatially-resolved stellar surface brightness profile that is corrected for dust would require realistic radiative transfer modeling (e.g., \citealt{camps15}) that incorporates the effects of scattering and extinction in the disk, and possible additional light contribution from the central AGN or star formation.  Such an exercise is beyond the scope of this paper.

If the dust extinction has been underestimated, then the amount of light included in the stellar surface brightness profile will be underestimated.  However, the amount of mass is set by the kinematics, which are not affected by dust extinction, so we would expect our best-fit models to overestimate $M/L$ if the dust extinction is underestimated. As a verification check, we compared our best-fit $M/L$ with $\sim$ 3500 galaxies in the MaNGA DynPop catalog \citep{zhu23}, each having a total stellar mass similar to MCG--06-30-15 ($\log M_{star}/M_{\odot}$ $\approx$ 10). Our best-fit $M/L$ of ($3.0\pm0.3$) $M_{\odot}/L_{\odot}$ is consistent with the typical 1$\sigma$ range of dynamical $M/L$ values for galaxies in this mass regime.

\begin{figure*}
    \centering
    \includegraphics[width=\linewidth]{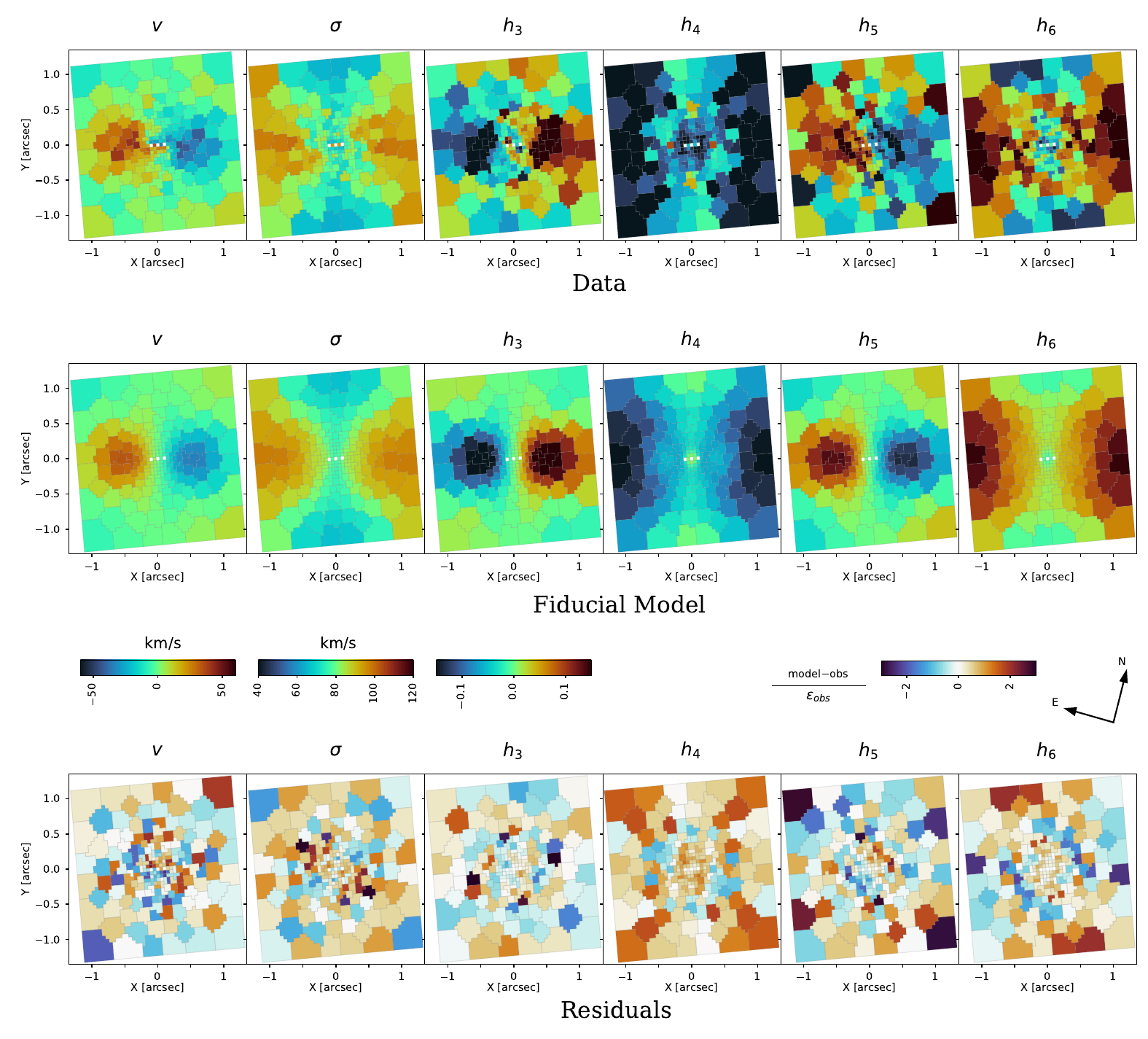}
    \caption{Comparison of kinematic maps derived from \texttt{pPXF} for the inner $3''\times3''$ of MCG--06-30-15. The top two rows show the velocity ($v$), velocity dispersion ($\sigma$), and higher order ($h_3$-$h_6$) moments of the Gauss--Hermite polynomials for the data (same as Figure \ref{fig:kinmap}) and the fiducial model ($q_d=0.2$, $q_b=0.6$, and $q_s=0.8$) respectively. The bottom row shows the residuals (= $\frac{model-obs}{\epsilon_{obs}}$) for each moment. All the bins from the full FOV are shown for completeness, though our models only included half of the kinematics because the other half are point symmetric. The orientation and the color scaling are the same as for Figure \ref{fig:kinmap}. The three white bins close to the center were severely affected by the AGN emission, and were masked out during the modeling. The highly negative $h_4$ values in the data may arise from the counter-rotating disk in the galaxy.}
    \label{fig:kin_comp}
\end{figure*}

\begin{figure*}
    \centering
    \includegraphics[width=0.8\textwidth]{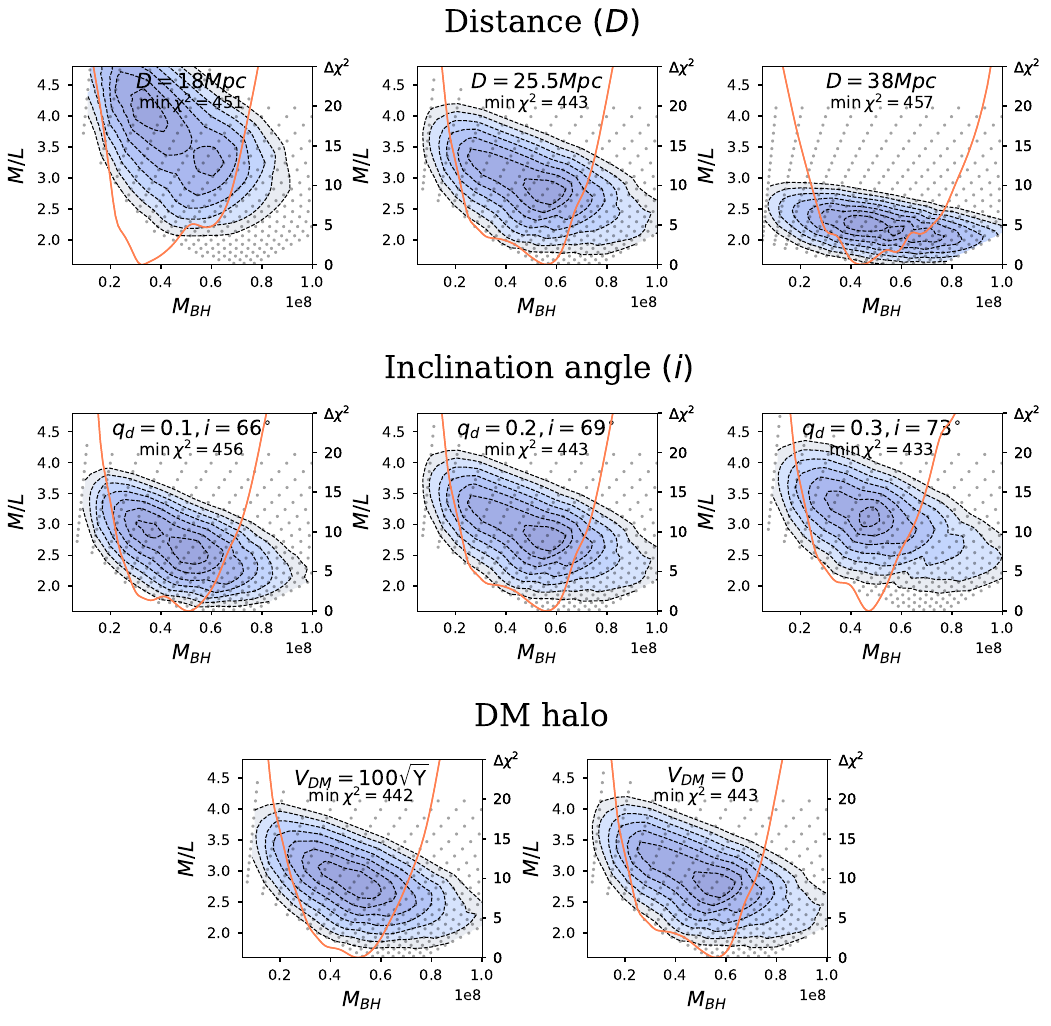}
    \caption{Comparison of $\Delta\chi^2$ for different parameter choices in the dynamical models. All the panels have the same flattening for the bulge and the stellar halo ($q_b=0.6$, and $q_s=0.8)$. The top and bottom rows have the same disk flattening ($q_d$=0.2) too. In the top row, we show how the adoption of different distances to the galaxy affects the $M_{BH}$ constraints. The distances shown are $D= 18$ Mpc, 25.5 Mpc (fiducial model), and 38 Mpc, from left to right. These distances are from the Extragalactic Distance Database and correspond to the two distances measured for galaxies in the same group as MCG--06-30-15 and their average. $M_{BH}$ and distance are linearly correlated, yielding a lower $M_{BH}$ for smaller distance and a higher $M_{BH}$ for larger distance. The middle row shows the effect of different disk flattening which correspond to different inclinations. The plotted disk flattenings are: 0.1, 0.2, and 0.3 corresponding to $i = $ 66$^{\circ}$, 69$^{\circ}$ (fiducial model), and 73$^{\circ}$, from left to right. The bottom row shows the contribution of the DM halo to our results. The left panel shows the model results when a DM halo is included, with a velocity of $100 \sqrt{\Upsilon}$ km/s, where $\Upsilon$ is the stellar mass-to-light ratio, and a scale radius $r_{DM} = 180''$, while the right panel shows the same models without a DM halo. The shape of the $\chi^2$ contours and the minimum $\chi^2$ value change slightly for models with/without DM.}
    \label{fig:chi2_dis}
\end{figure*}

\subsection{Black Hole Mass}
From our best-fit models, the black hole mass for MCG--06-30-15 is $M_{BH} = (4.4\pm1.4) \times 10^7 M_{\odot}$. This SD mass is at least a factor of 10 greater than the reported $M_{BH}$ from reverberation mapping of the AGN \citep{bentz16a}. However, there are a variety of issues that could affect either the SD mass that we present here or the RM mass.

On the reverberation mapping side, the reported mass is based on a measured time delay and velocity width for the broad H$\beta$ emission line, coupled with the adoption of a population-average scale factor.  This scale factor is, on average, of the right magnitude for reverberation mapping of local Seyferts, but may be inaccurate by a factor of $\lesssim10$ for any specific galaxy. The value adopted by \citet{bentz16a} for MCG--06-30-15 is $\langle f \rangle=4.3 \pm 1.1$ \citep{grier13}.  Assuming that most of the AGN-to-AGN difference in scale factor is due to their random inclination angles on the sky, a scale factor of $\langle f \rangle=4.3$ corresponds to a typical AGN inclination angle of $\sim 29$\degr. If the AGN in MCG--06-30-15 is actually viewed at a more face-on inclination ($i \lesssim$ 20\degr), then the black hole mass would be underestimated. Misalignments between large-scale galaxy inclination and AGN inclination are common, with a notable example being IC\,4329A with a broad-line AGN at an intermediate inclination hosted by an edge-on galaxy \citep{bentz23a}.

Modeling of velocity-resolved RM data allows the kinematics and the geometry (including the inclination) of the BLR to be constrained, and thus provides a direct constraint on $M_{BH}$ without the need to adopt a population-average scale factor \citep{cackett21}.  Comparisons of \mbh\ values from simple RM analysis and from velocity-resolved modeling show that the differences are generally within the typical factor of $2-3$ uncertainty that is often quoted for RM masses, except in the cases of the most face-on AGN where the differences may be as high as a factor of $\sim 10$ \citep{pancoast14b,villafana23}. 
While velocity-resolved modeling of the data presented by \citet{bentz16a} are only able to provide weak constraints, the results prefer a low black hole mass, $M_{BH}=1.9^{+5.7}_{-1.2} \times 10^6 M_{\odot}$, and a moderate BLR inclination of $i=40.0^{+29.5}_{-20.7}$\,deg (see Appendix~\ref{sec:velres_rm}).  Modeling of the broad Fe K$\alpha$ line also finds a moderate inclination ($i=33 \pm 3$\,deg) for the accretion disk  \citep{marinucci14}.  Thus, there is currently no likely candidate for the source of the discrepancy in the black hole masses from the reverberation side. A new reverberation mapping campaign targeting  MCG--06-30-15 (PI: Bentz) was carried out in 2024 and the observations are expected to enable a more thorough velocity-resolved reverberation analysis that will facilitate a stronger investigation into the assumptions inherent in the reverberation method.

There are potential issues on the stellar dynamical modeling side as well. Our best-fit $M_{BH}$ from SD modeling depends on the inclination angle of the galaxy disk. The surface brightness profile indicates an inclination of 69$^{\circ}$ for MCG--06-30-15. However, recovering the inclination angle from a 2D image of the galaxy always leaves room for uncertainty. We have explored models with different inclination angles and the results are shown in the middle row of Figure \ref{fig:chi2_dis}. It should be noted that the orbit-superposition method is intrinsically propensive towards edge-on systems, returning better fits for larger assumed inclination angles (see Section 5.3 in \citealt{thomas07}, \citealt{lipka21}). We note that $q_d$ = 0.3 ($i=73$\degr) yields a lower $\chi^2$ than $q_d$ = 0.2 ($i=69$\degr; our fiducial model). However,  disk galaxies with a $q_d \approx$ 0.3 are rare \citep{favaro25}. We also note that, while the min $\chi^2$ value changes, the best-fit values for $M_{BH}$ do not change significantly for models with higher inclination angle. Thus the uncertainty in the inclination angle has only a small effect on the dynamical $M_{BH}$ constraint.

The mathematical parameterization of the stellar kinematics is also a potential matter of concern. We note that $h_4$ is systematically less negative in our models than in the data (see Figure \ref{fig:kin_comp}). 
The low $h_4$ values in our data could be related to double-peaked velocity profiles: a line profile with negative $h_4$ has a flatter center and a more rectangular overall shape. While an infinite series of GH polynomials could fit any line shape, the limited S/N of astronomical spectra necessitates the use of a truncated series to avoid fitting the observational noise. With only a limited number of higher-order GH polynomial terms included, a double-peaked line profile may be approximated by a single-peaked shape with negative $h_4$.  This is a potential cause for concern in MCG--06-30-15 because \citet{raimundo13, raimundo17} found a counter-rotating disk within the central $\sim 4'' \times 4''$. The FOV of our data is small enough that the main rotational signature seen in the kinematics is the counter-rotating stellar population. But since the galaxy is highly inclined, the LOSVDs from most of the bins would include kinematic contributions from the larger-scale galactic disk, which is rotating in the opposite direction. We also note that the $h_4$ values are most negative at radii that are just larger than those associated with the maximum redshift/blueshift signatures of the counter-rotating disk, near the edges of our small field of view. Therefore, in the case of MCG--06-30-15, the negative $h_4$ could indicate that we are seeing kinematics from these nested and oppositely-rotating disks in many of the bins.

Models with lower $M_{BH}$ are mostly ruled out by large deviations in $h_4$ between the data and the model. Hence, if strongly negative $h_4$ values are a poor description of the actual LOSVDs, i.e., if GH parameterization is not the most appropriate choice for the kinematics in this system, {or if the values of $h_4$ returned by \texttt{pPXF} suffer from some unknown systematics (see, e.g., Section~3 in \citealt{mehrgan23} for discussion), then this issue would bias our measured SD mass towards higher black hole masses. This limitation in parameterization of the LOSVDs may be largely responsible for the discrepancy between the RM and SD masses measured for MCG--06-30-15. Finally, we also note that a negative $h_4$ can indicate that there is a bar present in the galaxy. \citet{brown13} have shown that in the case of barred galaxies, $M_{BH}$ may be overestimated when the bar is not modeled.

An exploration of the magnitude of the effect of the GH parameterization on the final $M_{BH}$ constraints is beyond the scope of this paper, but future studies that examine other parameterization choices will shed additional light on this issue. An alternative method is to use the histograms, or more generally, B-spline basis set \citep{merritt97, gebhardt00b, falconbarroso21,mehrgan23,gasymov24} for LOSVDs, but in that case finding an optimized smoothing factor is important because oversmoothing will return a pure Gaussian function. Moreover, in this case, the errors in different velocity moments are not independent, and their full correlation matrix should ideally be incorporated into the models. Another potential technique was explored by \citet{houghton06}, who used `eigen velocity profiles' to represent the non-Gaussian LOSVDs in NGC 1399 which has a decoupled core. However, it remains to be seen if this technique could be used for modeling galaxies like MCG--06-30-15.

To constrain the larger-scale stellar kinematics, and have a better understanding of the wider-field stellar population and the DM halo, it is generally recommended to also include kinematic measurements from a larger FOV. \citet{raimundo17} obtained $H$-band and $K$-band spectroscopic data from SINFONI covering the central $8''\times8''$ to investigate the AGN fueling in MCG--06-30-15. They found $M_{BH} \le 6\times 10^7 M_{\odot}$ from Jeans Anisotropic Modeling method, which is consistent with our measurement of the black hole mass using the Schwarzschild orbit-superposition method. We explored the possibility of including these wider FOV observations in our modeling. However, due to the limited S/N of the dataset, we were not able to construct models that provided a reliable constraint for $M_{BH}$, and ultimately, we did not include these data in our analysis. 

Lastly, the $M_{BH}$ constraints from our modeling also linearly depend on the distance to the galaxy. Unfortunately, there is no reliable distance to MCG--06-30-15 to date. Based on the Extragalactic Distance Database (EDD)\footnote{EDD can be accessed at \url{ https://edd.ifa.hawaii.edu}} by \citet{tully09}, the average distance to the galaxy group of MCG--06-30-15 is $25.5\pm 3.5$ Mpc \citep{tully13}, which is the value we adopt in this work.  However, only two out of five members in this group of galaxies have their distances measured from methods other than their redshift, and the distances are $D=18$ Mpc and $D=38$ Mpc. This large range of potential distances for MCG--06-30-15 changes our models quite significantly, and thereby our constraint on $M_{BH}$. The first row in Figure \ref{fig:chi2_dis} shows how the black hole mass and the $M/L$ values depend on the assumed distance to the galaxy. The dynamical $M_{BH}$ is directly proportional to the distance to the source while the mass-to-light ratio is inversely proportional. While the distance is currently not well constrained, a much smaller distance would be required to completely account for the discrepancy between the SD mass we present here and the RM mass. The uncertainty on the distance could reasonably be expected to impact the $M_{BH}$ measurement by up to a factor of $\sim2$. Additionally, we note that RM is based on light-travel time, which constrains physical distances rather than angular distances, and therefore does not depend on the adoption of a specific value for the distance to the source.

Efforts to more accurately constrain the distance to this galaxy are underway. The redshift, $z = 0.00775 $ (corresponding luminosity distance, $D_L = 32.5 $ Mpc) is too low to provide a reliable distance using Hubble's law because the recession velocity of this galaxy may be strongly affected by gravitational interactions with its neighbors \citep{tully08, tully13}. There is no major star-forming region in MCG--06-30-15, making it a poor candidate for a Cepheid-based distance measurement. It is also not likely to be within the distance limit ($D \leq 20$  Mpc, \citealt{sakai96, freedman19}) for the tip of the red giant branch method. The strong dust lane also eliminates the possibility of using the surface brightness fluctuations technique \citep{tonry97}.

One technique that may be appropriate for measuring an accurate distance to MCG--06-30-15 is the \citet{tully77} technique.
\citet{robinson19} detected a tentative \ion{H}{1} 21\,cm signal at the $3\sigma$ level using the 100m Green Bank Telescope (GBT), but the S/N was too low to constrain a distance. We are following up with deeper GBT spectroscopy of this galaxy to better characterize the integrated \ion{H}{1} 21\,cm emission from the galaxy with the intent of determining a stronger distance constraint for MCG--06-30-15, and therefore, a more accurate black hole mass.

In conclusion, we have identified a few different  sources that may contribute to the discrepancy of $M_{BH}$ in MCG--06-30-15 from RM and SD modeling. It is not yet clear if any one of these factors alone is sufficient to overcome the overall discrepancy between the two mass measurement methods, and the magnitude of the effect has not been quantified for all of these factors.  A more accurate distance may shift the dynamical mass by up to 50\% whereas the uncertainty from the inclination angle has only a modest effect.  However, the magnitude of the effect on the black hole mass resulting from a poor parameterization of the underlying stellar kinematics, which we suspect could be an issue for this galaxy, has not yet been investigated. Further work is necessary to accurately quantify the relative contributions from these different sources of uncertainty and potentially resolve the discrepancy in the black hole mass measurements for this galaxy.

\section{Summary}\label{sec:summary}

We present the first black hole mass from stellar dynamical modeling in MCG--06-30-15 using the Schwarzschild orbit-superposition technique. For this work, we have acquired the highest resolution AO-assisted $K$-band spatially resolved spectroscopy of this galaxy using SINFONI on the VLT. The stellar kinematics were derived by fitting the CO absorption bands in the spectra, and the LOSVDs were parameterized by Gauss--Hermite polynomials. A three-component surface brightness profile of the galaxy was derived by fitting an \textit{HST} medium-bandpass $V$ image. We used the open-source code FORSTAND to construct dynamical models that allowed the investigation of different shapes for the components of the galaxy. We constructed and compared dynamical models exploring a range of values for different parameters, i.e., stellar $M/L$, distance, flattening of galaxy components, etc. From the best-fit models, we have determined the black hole mass $M_{BH}$ = $(4.4\pm1.4) \times 10^7 M_{\odot}$ and a $V$-band $M/L$ of $(3.0\pm0.3)$ $M_{\odot}/L_{\odot}$ for MCG--06-30-15 within 1$\sigma$ confidence level. This $M_{BH}$ constraint for MCG--06-30-15 is consistent with the upper limit suggested by \citet{raimundo17}. However, our measurement of $M_{BH}$ is at least a factor of 10 greater than the reported black hole mass from RM \citep{bentz16a}. 

The discrepancy between the masses from these two different methods may partially arise from the uncertainty in the distance to the galaxy. Furthermore, the dynamical mass is potentially biased high because of the presence of a counter-rotating disk in MCG--06-30-15, which may not be well described by the typical method of parameterizing stellar kinematics with Gauss--Hermite polynomials. Future efforts to more accurately extract the LOSVDs from the spectra using different mathematical parameterizations will help assess the magnitude of this effect on the reported mass. A more reliable distance measurement and upcoming velocity-resolved RM analysis for MCG--06-30-15 will also help to shed light on the differences in the reported masses for this galaxy.

\begin{acknowledgements}
We thank the anonymous referee for suggestions that improved the presentation of this work.
N.\ Das and M.C.\ Bentz gratefully acknowledge support from the NSF through grants AST-2009230 and AST-2407802.
E.\ Vasiliev acknowledges support from an STFC Ernest Rutherford fellowship (ST/X004066/1).
S.I.\ Raimundo acknowledges the support from STFC via grant reference ST/Y002644/1.
M.\ Vestergaard acknowledges support from the Independent Research Fund Denmark (grants DFF 8021-00130 and  3103-00146) and by Carlsberg Foundation grant CF23-0417.
Based on observations collected at the European Southern Observatory under ESO program 0102.B-0362(A).
IRAF was distributed by the National Optical Astronomy Observatory, which was managed by the Association of Universities for Research in Astronomy (AURA) under a cooperative agreement with the National Science Foundation. N.\ Das thanks Dr. Katherine Merrell for insightful suggestions and helpful discussions.

\end{acknowledgements}

\begin{contribution}
N.\ Das carried out the kinematic analysis, the modeling, and wrote the manuscript.  M.C.\ Bentz reduced the observations, conducted the \texttt{Galfit} analysis and velocity-resolved reverberation analysis, and oversaw the project. E.\ Vasiliev provided assistance and expertise with the modeling. M.\ Valluri and C.A.\ Onken provided scientific expertise and assisted with the interpretation. S.I.\ Raimundo and M.\ Vestergaard were instrumental in helping obtain the SINFONI observations and provided supplemental observations.  All authors contributed to the presentation of the results.
\end{contribution}

\facilities{VLT:Melipal, HST:WFC3}

\software{Molecfit \citep{smette15,kausch15}, pPXF \citep{cappellari04, cappellari17}, Galfit \citep{peng02,peng10}, FORSTAND \citep{vasiliev20}}

\appendix

\section{Velocity-Resolved Reverberation Mapping Analysis}\label{sec:velres_rm}

Forward modeling of the velocity-resolved reverberation response in spectroscopic monitoring data is one method for providing a more accurate reverberation-based mass. This method is analogous to dynamical modeling of stellar or gas kinematics, relying on a framework of self-consistent models working in time delay and velocity space to explore possible geometries and kinematics of the broad-line emitting gas that is deep within the potential well of an actively accreting supermassive black hole. Velocity-resolved modeling of reverberation-mapping data directly constrains $M_{BH}$ along with the detailed BLR motions and structure, thus avoiding the adoption of a scale factor.

Following the procedures outlined in detail by \citet{bentz21c,bentz23b}, we used the velocity-resolved modeling code {\tt CARAMEL} \citep{pancoast11,pancoast14a} to investigate the potential for the existing reverberation-mapping data for MCG--06-30-15 to provide additional insight into the details of the BLR, and thus the black hole mass.  The models were able to recover the typical time delay --- $\tau_{mean}=6.63^{+2.10}_{-2.41}$\,days compared to $5.3\pm1.8$\,days as reported by \citet{bentz16a} --- but the limited time duration of the monitoring campaign and relatively low amplitude of variations in the dataset resulted in only modest constraints for many other parameters. 

In Figure~\ref{fig:posteriors}, we show the posterior probability distributions for $M_{BH}$, $\tau_{mean}$, and the broad-line region inclination, $\theta_i$.  We note that even with the modest constraints available, the models disfavor black hole masses $\gtrsim 10^7 M_{\odot}$ as well as broad-line inclinations $\lesssim 15^{\circ}$.

\begin{figure}[h!]
\centering
\includegraphics[width=3.0in]{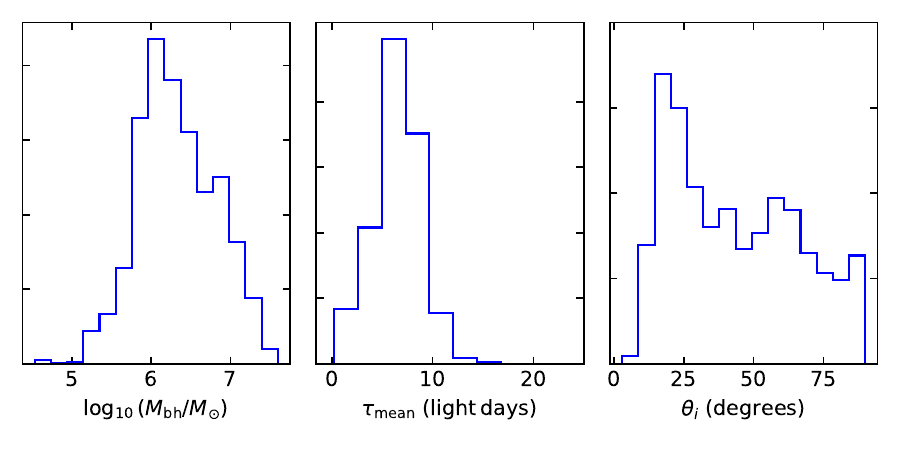}
\caption{Posterior probability distributions for several parameters of the velocity-resolved reverberation models: $M_{BH}$, mean time delay $\tau_{mean}$, and BLR inclination to our line of sight, $\theta_i$.}\label{fig:posteriors}
\end{figure}

\bibliography{ND_ref}{}
\bibliographystyle{aasjournal}

\end{document}